\documentclass[%
 aip,
 amsmath,amssymb,
 reprint,%
]{revtex4-1}

\usepackage{graphicx}
\usepackage{dcolumn}
\usepackage{bm}
\usepackage{xcolor}

\usepackage[utf8]{inputenc}
\usepackage[T1]{fontenc}
\usepackage{mathptmx}
\usepackage{etoolbox}

\makeatletter
\def\@email#1#2{%
 \endgroup
 \patchcmd{\titleblock@produce}
  {\frontmatter@RRAPformat}
  {\frontmatter@RRAPformat{\produce@RRAP{*#1\href{mailto:#2}{#2}}}\frontmatter@RRAPformat}
  {}{}
}%
\makeatother
\begin{document}


\title[]{Discrete-space and -time analogue of a super-diffusive fractional Brownian motion}
\author{Enzo Marinari}
 \email{enzo.marinari@uniroma1.it}
 \affiliation{Dipartimento di Fisica, Sapienza Universit\`{a} di Roma, P.le A. Moro 2, I-00185, Roma, Italy 
}
\affiliation{INFN, Sezione di Roma 1 and Nanotech-CNR, UOS di Roma, P.le A. Moro 2, I-00185, Roma, Italy
}
\author{Gleb Oshanin}
\affiliation{Sorbonne Universit\'e, CNRS, Laboratoire de Physique Th\'eorique de la Mati\`{e}re Condens\'ee (UMR 7600), 4 Place Jussieu, 75252, Paris, Cedex 05, France
}

\date{\today}

\begin{abstract}
We discuss how to construct reliably well "a lattice and an integer
time" version of a super-diffusive continuous-space and -time
fractional Brownian motion (fBm) -- an experimentally-relevant
non-Markovian Gaussian stochastic process with an everlasting
power-law memory on the time-evolution of thermal noises extending
over the entire past. We propose two algorithms, which are both
validated by extensive numerical simulations showing that the ensuing
lattice random walks have not only the same power-law covariance
function as the standard fBm, but also individual trajectories follow
those of the super-diffusive fBm.  Finding a lattice and an integer
time analogue of a sub-diffusion fBm, which is an anti-persistent
process, remains a challenging open problem.  Our results also clarify
the relevant difference between sub-diffusive and super-diffusive fBm,
that are frequently seen as two very analogous realizations of
processes with memory. They are indeed substantially different.
\end{abstract}

\maketitle

\begin{quotation}
In the experimentally-relevant non-Markovian Gaussian stochastic
process of fractional Brownian motion (fBm), the behavior of a
particle's trajectory on a positive time interval is dominated by a
scale-free memory of the thermal noise acting on the particle over an
infinitely-long time interval prior to the time origin.  To properly
define a lattice counterpart of such a process, i.e., a process with
strictly integer increments stepping at integer time instants, but
which possesses exactly the same covariance function as the original
continuous-space and -time fBm, represents quite a non-trivial problem
and amounts to devising a special "coin": this should be "fair", but
only on average over all possible sequences of consecutive tossing
events, and should have an infinite memory such that within each given
sequence of possible outcomes the increments will be strongly
correlated with a pre-defined correlations' structure. The problem of
determining such a "coin" for a super-diffusive fBm is discussed in
the present paper.
 \end{quotation}

\section{\label{sec:level1}Introduction}

Experimental analysis of particles' dynamics in complex environments
often reveals significant departures from the standard Brownian
motion, showing instead a power-law behavior of the form
$\langle X^2(t)\rangle\sim t^{\alpha}$, where $\langle X^2(t)\rangle$ is the
mean square displacement of the particle from its initial position at
time instant $t$, the angle brackets here and henceforth denote
averaging over a statistical ensemble of trajectories, and $\alpha
\neq 1$ is the so-called \textit{anomalous} diffusion exponent
\cite{a,b,bb,c}.  Such a law of "anomalous diffusion" (AD), with
$\alpha < 1$ corresponding to sub-diffusive processes and $\alpha > 1$
to super-diffusive ones, is observed in a variety of different
chemical, physical and biophysical systems.  Depending on the case at
hand, the AD may occur due to a variety of different factors -- the
presence of a random potential \cite{aa,comtet} or of random flow
patterns \cite{blum,enzo,nech}, a temporal pausing at some positions
characterized by a broad distribution of pausing times
\cite{ralf1,ralf2}, complexity of the particle itself (e.g., it can be
a polymer with a complicated internal dynamics affecting the
translational diffusion) \cite{doi1,doi2}, molecular crowding effects
\cite{denis,olivier}, effects of correlations between the increments
(see, e.g., Ref. \onlinecite{comtet}), and more.  An interesting
approach \cite{bender24} considers the motion of a test particle
surrounded by $N$ Brownian particles with different masses,and shows
that in some conditions the test particle can undergo fBm.  More
discussion about the origins of the AD can be found in
Ref. \onlinecite{ralf6}.

There exists a plethora of very diverse mathematical models of AD that
emphasize different factors and underlying mechanisms. Some of these
models exist in both continuous- and in the discrete-space (lattice)
formulations \cite{hughes}, where the former are sometimes more
flexible and are amenable more readily for an analytical description,
while the latter are typically more difficult to tackle analytically
but clearly appear more suitable for a numerical analysis.  It is
therefore very advantageous to know both formulations, which permits
to properly discretize a continuous-space formulation when a numerical
verification is needed or, conversely, to properly define a
continuous-space limit for a given discrete-space model.  In
particular, both continuous- and discrete-space versions of the
celebrated Sinai model of a logarithmically-confined sub-diffusion in
a random forcing landscape are well-known \cite{sinai} (see also
Refs. \onlinecite{aa,comtet,enzo2}).
In turn, the continuous-space
models of a sub-diffusive random motion due to a temporal trapping or
of a super-diffusion due to long-range relocations, described by the
so-called fractional Fokker-Planck equations, can be analyzed with
a very direct approach \cite{ralf7,ralf8}.
The standard
simulation algorithms for latter models make use of a memory-less fair
"coin" which defines the jump direction, where the term "fair" means
that the coin, when tossed at each trial should have an equal chance
of landing either side up (we denote a face up side by $+1$ and the
face down side by $-1$ in what follows) independently of the outcomes
of the previous trials. Then, one uses some \textit{a priori} known
probability distributions of either the time-intervals between the
consecutive jumps, or of the jump lengths.  In this way, one simulates
efficiently the continuous-space AD and determines various properties
of interest. Clearly only continuous walks can be simulated in this
way. 

Concurrently, such algorithm for continuous walks cannot be realized for many naturally-occurring processes with a memory. 
For instance, it is not applicable for the experimentally-relevant \cite{krapf1,krapf2} non-Markovian Gaussian 
process of the so-called \textit{fractional} Brownian motion (fBm)  studied  in depth by Mandelbrot and van Ness \cite{mvn} (see Refs. \onlinecite{baruch1,baruch2} for a more ample discussion). 
A salient feature of such a process is that here the behavior of a particle's random trajectory on a positive time interval is dominated by its behavior during an infinite time interval prior to the time origin (see below). 
Framing the fBm in terms of an effective Langevin equation, one has that 
here the increment $\delta_t = dX/dt$ is proportional to a random force which is a \textit{Gaussian} noise with long-ranged (either positive or negative) power-law temporal correlations, which extend over the entire 
past -- the so-called \textit{fractional} Gaussian noise (see Refs. \onlinecite{mvn} and \onlinecite{baruch1,baruch2,ralf11}). 
As a physical example, one may consider the dynamics of a tagged bead in an infinitely long Rouse polymer chain\cite{doi1,doi2};  here, 
the trajectories of the bead will correspond to 
realizations of the fBm process with a particular value of $\alpha$ for positive observation times,
 if the entire chain is let first to thermalize itself over an infinite time-interval prior the observation begins. 

To properly define a lattice counterpart of such a process, i.e., a  process with an everlasting scale-free
memory which has strictly integer increments, represents therefore quite a non-trivial problem 
and amounts to 
devising a special "coin". This coin should be "fair", but only on average over all possible sequences of consecutive tossing events, and should have an infinite memory such that within each given sequence  $\{\pm 1,\pm  1, \ldots, \pm 1\}$ of possible outcomes the increments will be strongly correlated 
 with a controllable, pre-defined correlations' structure. We note parenthetically that such a coin 
can also be useful for a variety of other problems, e.g., for
 studying storage/retrieval properties of the emblematic Hopfield model \cite{amit}, 
 which has been amply analyzed in the past, but solely for the binary sequences of \textit{uncorrelated} variables.  
 Many other examples of naturally-appearing correlated 
 binary sequences and the ways to generate them can be found in \cite{1995_HAVMAN}.
 
 In this paper we discuss this problem of generating binary sequences
 $\{\pm 1,\pm  1, \ldots, \pm 1\}$ of increments with positive power-law intra-sequence correlations, which can be considered as an integer, two value analogue of fractional Gaussian noise providing a discrete-space and -time version of a
super-diffusive fractional Brownian motion.
 The paper is outlined as follows. In Sec. \ref{sec:2} we present a brief reminder on fBm processes in continuous space and time. In Sec. \ref{sec:3} we discuss two approaches permitting us to construct a lattice version of the fBm, which are both validated numerically in Sec.   \ref{sec:4}. We conclude in Sec. \ref{sec:5} with a brief recapitulation of our results.

\section{Standard fractional Brownian motion in continuous-space and -time}
\label{sec:2}

Here we briefly recall the standard definition of fBm -- in fact, a
family of centered one dimensional anomalous stochastic processes with
everlasting power-law correlations between the increments. We
concentrate here on the so-called one sided Mandelbrot - van Ness fBm
for which the time variable $t$ is defined on the interval
$(0,\infty)$ but the behavior of the trajectory $X(t) $ for $t > 0$
depends explicitly on the evolution of the noise over the entire
past. More specifically, $X(t)$ (with $X(t=0) = 0$) is defined as
\begin{align}
\label{a}
&X(t)  = \frac{1}{\Gamma(H+1/2)} \Bigg\{\int^{t}_0 \left(t - \tau\right)^{H-/2}  \, \zeta_{\tau} \, d\tau \nonumber\\
&+ \int^0_{-\infty} \Big[(t-\tau)^{H-1/2} - (-\tau)^{H-1/2}\Big] \, \zeta_{\tau} \, d\tau \Bigg\} \,,
\end{align}
where $\Gamma(z)$ is the Gamma function,  $\zeta_{\tau}$ is a  zero mean Gaussian white noise with the covariance function
\begin{align}
\langle \zeta_{\tau} \zeta_{\tau'} \rangle = \delta(\tau - \tau') \,,
\end{align}
with $\delta(\tau)$ being the delta function and the angle brackets,
as above, denoting averaging over all possible realizations of the
noise. Lastly, $H$ is a crucial parameter - the so-called Hurst
exponent \cite{mvn} - $H \in (0,1)$.

After some calculations, one finds that the covariance function of the
process defined in Eq. \eqref{a} obeys
\begin{align}
\label{b}
\left \langle X(t) X(t') \right \rangle = \frac{1}{2} \left(t^{2H} + t'^{2H} - |t - t'|^{2H}\right) \,.
\end{align}
The following important comments on the result in Eq. \eqref{b} are in order:\\
-- Setting $t = t'$ one finds that the mean squared displacement $\langle X(t) X(t') \rangle = t^{2 H}$, such that the above mentioned anomalous diffusion exponent  $\alpha = 2H$. Consequently, for $H < 1/2$ the process in Eq. \eqref{a} is sub-diffusive and for $H > 1/2$ is super-diffusive. For $H=1/2$, as one may infer directly from Eq. \eqref{a}, the second term becomes identically equal to zero, so that the process ceases to depend on the past, while the first term is just an integral over the white noise. In consequence, in the borderline case $H = 1/2$ the process $X(t)$ is a Markov process without memory and is just the standard Brownian motion with delta correlated increments.\\
-- Differentiating the expression \eqref{b} over $t$ and $t'$, one has
\begin{align}
\label{c}
C(|t-t'|) = \left \langle \delta_t \delta_{t'} \right \rangle = \left \langle \frac{dX(t)}{d t} \frac{d X(t')}{dt'} \right \rangle = \frac{H(2H - 1)}{|t - t'|^{2 - 2 H}} \,.
\end{align}
Therefore, the increments $\delta_t$ of the one sided fBm are stationary, since their covariance function depends only on the difference $t-t'$. Further on, 
correlations between the increments vanish for $H = 1/2$, as they should, are positive for $H > 1/2$ and negative for $H < 1/2$. Hence, a  super-diffusive motion for $H > 1/2$ is realized, as can be expected on intuitive grounds,  due to a certain \textit{persistence} of random motion - if an increment on a previous move was positive, it will be most likely positive on the next one too. Conversely, dynamics for $H < 1/2$ is \textit{anti-persistent}; that being,  it is most likely that the increments on successive moves change their signs. Expression \eqref{c} also shows that the decay of correlations proceeds in a very different way for sub- and super-diffusions. For sub-diffusion the exponent $z = 2 - 2H$ characterizing the decay of correlations with $d\tau = |t - t'|$ exceeds unity, which signifies that correlations are integrable and are not sufficiently long-ranged indeed to ensure alone the sub-diffusive motion. This rather takes place due to some intricate interplay of the sign of increments on successive moves. On contrary, for $H > 1/2$ the exponent $z$ is less than unity such that the correlations are non integrable. In this case, this is precisely this long-ranged persistence which entails the super-diffusive motion. We will exploit this circumstance in what follows.

We close this Section by emphasizing that $X(t)$ in Eq. \eqref{a} is
just a particular member of the family of fBms: there exist two
alternative definitions bearing the same name which differ mainly by
the interval on which the time variable is defined, and hence, by the
form of the corresponding fractional integral. These are the
Riemann-Liouville form with $t \in (0,T)$ introduced by L\'evy
\cite{levy}, and the so-called two sided fBm with $t \in
(-\infty,\infty)$, put forth by Kolmogorov \cite{kol} and studied in a
great detail by Mandelbrot and van Ness \cite{mvn}.  A recent
comparison of different versions of the fBm as well as explicit
expression for the measure of their individual trajectories can be
found in Refs. \onlinecite{baruch1,baruch2}.  An extensive discussion
of various aspects of the fBm is presented in
Refs. \onlinecite{2012_NOURDIN_BOOK_FBM,2018_EVAKAM_BOOK_FBM,2023_EVAKAM_BOOK_FBM}.

\section{Discrete-space and -time analogue of a super-diffusive fBm}
\label{sec:3}

As stated above, our main goal here is to define an algorithm 
that permits 
 to generate a super-diffusive random walk which evolves on the integer number line in a discrete time, 
 with \textit{a priori} given Hurst index $H > 1/2$ and the covariance function of the increments that obeys Eq. \eqref{c}. 
 In other words, we seek a "fair",  on average, coin with long-ranged correlations which allows to generate
 a discrete-space and -time analogue of a super-diffusive fBm. 
 To this end, we consider two different approaches which both rely first on constructing 
 a continuous time trajectory $X(t)$ in Eq. \eqref{a} using an algorithm described in the next Section \ref{sec:4}, and then, on an appropriate discretization of the  increments. \\

-- In the first scenario,  which we call a "trivial level discretization" (TLD),
we determine the increments $\delta_t$ of the continuous time trajectory $X(t)$ in Eq. \eqref{a} recorded at integer time moments $t$  and define new increments $\triangle_t$ according to the following very simple rule: $\triangle_t = - 1$ if $\delta_t < 0$ and $\triangle_t  = + 1$, otherwise.  A similar approach has been previously used in Ref. \onlinecite{1995_HAVMAN} but without a detailed analysis of the properties of the ensuing correlated nearest neighbor random walk, 
which steps at each tick of the clock on a lattice of integers.  \\

-- In the second scenario, we put forth a slightly more sophisticated "first order discretization" (FOD) scheme based on
an  appropriately discretized Langevin equation with the fractional Gaussian noise. 
We start again with the original continuous-space and -time trajectory $X(t)$ in Eq. \eqref{a}, determine the corresponding increments $\delta_t$, and then introduce an integer valued positive definite scale factor $\varphi_t$ of the form 
\begin{equation}
\varphi_t\;=\; \left\lfloor \left|\delta_t\right| + 1/2\right\rfloor, 
\end{equation}
where $\lfloor \ldots\rfloor$ is the floor function, i.e., the function that takes as input a real number $x$ and maps it to the greatest integer less than or equal to $x$.
Consequently, 
$\varphi_t=0$ if the original increment $|\delta_t| < 1/2$, $\varphi_t=1$ if $1/2 < |\delta_t| < 1$, and etc.  
Next, we stipulate the increment $\triangle_t$ of the  
discrete-space and -time random walk generated within the FOD scenario  
obeys $\triangle_t = \pm \varphi_t$, where the sign is the same as of the increment $\delta_t$. 

Because such a random walk pauses at its position when occasionally $|\delta_t| < 1/2$, 
 the time $T_{{FOD}}$ needs to be rescaled to permit for a direct comparison of the ensuing trajectories with the original continuous-space and -time trajectories and the trajectories obtained within the TLD binarization. To this end, we renormalize the individual time step $\tau_{{FOD}}'$ of the random walk by setting
\begin{equation}
\tau_{{FOD}}' = \tau \times T /  T_{{FOD}}\; ,
\end{equation}
where $\tau$ is the incremental time of the trajectory $X(t)$ and $T$
is the total observation time.

Both our discretization procedures only take an order of $T$
computational steps, that implies that the complexity of the combined
algorithm is the same of the original algorithm that generated the
fBm.
 
\section{Numerical simulations}
\label{sec:4}

The fact that the fBm with the Hurst index $H \ne\frac12$ is a non-Markovian process for which 
the sign and the value of an increment at each consecutive step 
depend, in virtue of Eq. \eqref{a}, on the  
behavior in the entire past renders its numerical analysis quite difficult.  
In our numerical simulations we will follow here the algorithm elaborated by Dieker 
(see his PhD Thesis \cite{2004_DIEKER} and Ref. \onlinecite{2003_DIEKER_MANDJES} for more details)
and namely, the one based on the so-called Davies-Harte or circulant method~\cite{1987_DAVHAR,1994_WOOCHA,1997_DIENEW,1999_CHAWOO}. 
Note that this algorithm appears to be very efficient;  
indeed, when applied to a system with $N$ nodes the simulation time scales only like $N\ln(N)$, as compared to other methods for which the time scales like $N^2$ (this happens for example with the Hosking method \cite{Hosking1984}) or even as $N^3$ (see Ref. \onlinecite{2004_DIEKER}). This allows
us to extend our numerical analysis to very large times, inaccessible via other approaches.
The Davies-Harte or circulant method is based on the use of the Fast Fourier Transform. The Hosking method, that is simpler in implementation but dramatically slower for large values of $N$, is based on a recursion.
 In what follows, we use the circulant method to 
generate trajectories $X(t)$ consisting of $T=2^{22}$ steps and choose $H = 0.8$ as 
a representative example of a super-diffusive fBm. In doing so, we construct the discrete-space and -time analogues of a super-diffusive fBm using both above mentioned scenarios. 

In the case of super-diffusion we study a number of values of $H$
ranging from 0.6 to 1.0. We find a very consistent picture. Without
loss of generality we mainly report here about the case $H=0.8$, that
is very similar to the other cases. On the contrary our method does
not work for sub-diffusion, making clear a substantial difference among
the two regimes. Superdiffusion is universal, while in subdiffusion
non universal contributions play a role. 

\begin{figure}
    \centering
    \includegraphics[width=8cm]{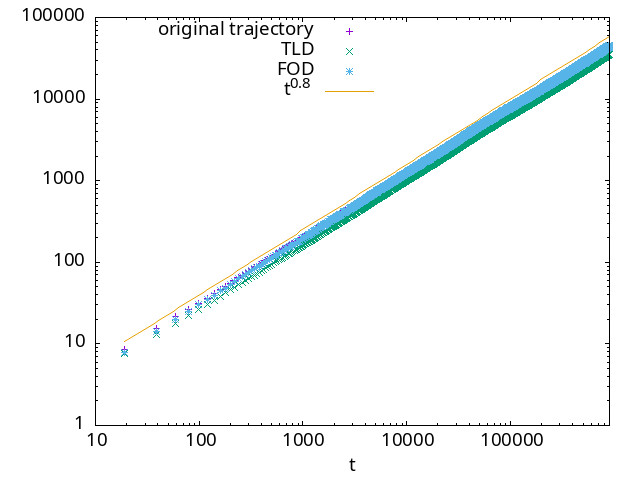}
    \caption{Ensemble averaged absolute value of $X(t)$ as function of
      time for the continuous-space and -time fBm with the Hurst index
      $H = 0.8$ (pluses) and its discrete-space and -time counterparts
      evaluated within the TLD (crosses) and the FOD (asterisks)
      binarization schemes. The magenta curve is a guide to the eye
      indicating the slope $t^{0.8}$.  Averaging is performed over $2
      \times 10^3$ realizations of the processes.
      Error bars are smaller than the size of the points.}
    \label{FIG_average_h0208}
\end{figure}

We start with the comparison of ensemble averaged (over $2 \times
10^3$ realizations) properties.  In Fig. \ref{FIG_average_h0208} we
plot the ensemble averaged absolute value of $X(t)$ as function of
time, as well as the corresponding results of the TLD (crosses) and
the FOD (asterisks) binarization schemes. We observe a perfect
agreement between the three sets of discrete data noticing, however,
that the more sophisticated FOD approach seems to work better than the
simplest TLD one - for the former the corresponding data appears
closer to the result for the standard fBm than for the latter one.  It
is useful to also compare numerically an averaged observable computed
on the original and on the improved trajectory. As an example we
compute the average of $X(t)^2$ for $t=2^{12}$ by considering
20000 trajectories, and we find $5.535(1) 10^6$ for the original
trajectory and $5.537(1) 10^6$ for the the FOD trajectory.

 \begin{figure}
    \centering
    \includegraphics[width=8cm]{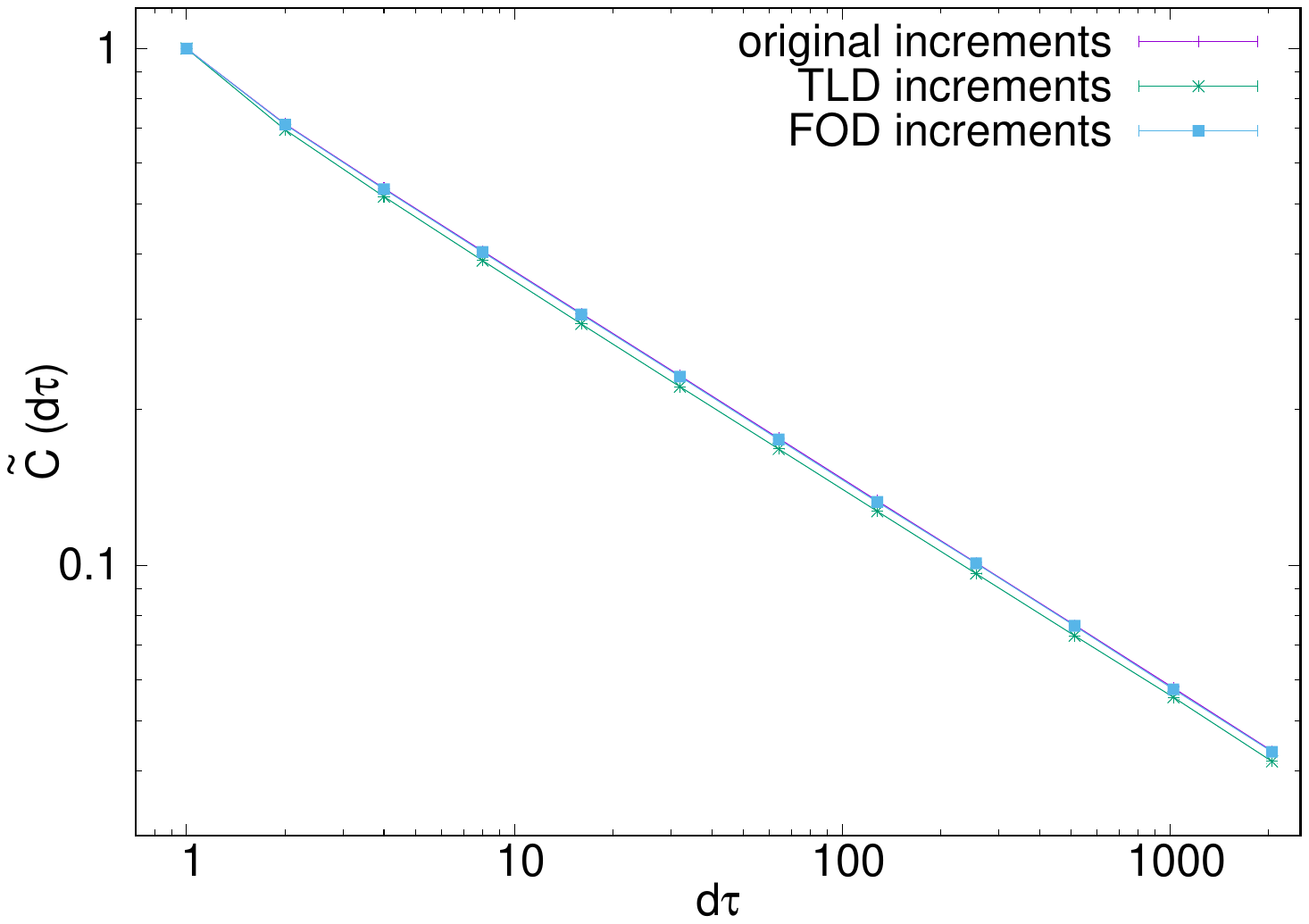}
    \caption{Log-log plot of the reduced covariance function $\tilde{C}(d\tau)$ defined in Eq. \eqref{EQ_CF1} as function of $d\tau = |t - t'|$ for $H = 0.8$. The line with ticks is the result for the continuous-space and -time process in Eq. \eqref{a}, crosses correspond to the results obtained within the TLD scheme, while the filled squares - to the FOD scheme. Averaging is performed over $2 \times 10^3$ realizations of the processes.      Error bars are smaller than the size of the points.
}
    \label{fig:2}
\end{figure}

Remember next the key fact about Gaussian processes (e.g., the fBm under study) that they can be completely defined by their second order statistics. If a Gaussian process is centered, i.e., its first moment is identically equal to zero,  the form of the covariance function of the process or, equivalently, of its increments, completely defines the process' behavior. For the continuous-space and -time process in Eq. \eqref{a} the covariance function of the increments obeys Eq. \eqref{c}. Correspondingly, our next aim is to  check numerically whether the covariance functions of the increments of processes obtained via the TLD and FOD schemes follow exactly the same power-law form as in Eq. \eqref{c}.
 For convenience, we will study 
not the property defined in expression \eqref{c}, but rather its reduced form $\tilde{C}(d\tau)$ which follows
\begin{equation}
\tilde{C}(d\tau) \equiv \langle\delta_t\delta_{t+d\tau}\rangle/\langle\delta_t\delta_{t+1}\rangle, 
    \label{EQ_CF1}
\end{equation}
i.e., $\tilde{C}(d\tau)$ is the covariance function of the increments divided by its value when the two times, $t$ and $t'$, differ by unity. Therefore,  
by construction, the results for the continuous-space and -time process $X(t)$ should coincide with the ones for the TLD and FOD schemes when $d\tau=1$. For the continuous-space and -time process this function is evidently given by
\begin{equation}
\tilde{C}(d\tau) = \frac{C(|t-t'|)}{H(2H-1)} = \frac{1}{(d\tau)^{2 - 2 H}} \,,
\end{equation}
where $C(|t-t'|)$ is defined in Eq. \eqref{c}.
 
In Fig. \ref{fig:2} we depict  $\tilde{C}(d\tau)$ as function of $d\tau = |t - t'|$ for $H=0.8$ for the original continuous-space and -time process $X(t)$ (line with ticks) and its discrete-space and -time counterparts obtained within the TLD scheme (line with asterisks) and the FOD (line with filled squares) schemes. We observe a perfect agreement between the three lines for the variable $d\tau$ which spans more than three decades, noticing that, again, the FOD scheme works somewhat better than the TLD one, being almost indistinguishable from the covariance of the parental fBm process.  

\begin{table}
\caption{\label{Table} The exponent $\epsilon$ characterizing the
  power-law decay of the increment-increment correlation function in
  Eq. \eqref{fit} with $d\tau = |t - t'|$.  The Table presents the
  values of $\epsilon$ for the parental continuous-space and -time
  process $X(t)$ ($\epsilon_{\rm X}$) and its discrete-space and -time
  counterparts obtained within the TLD ($\epsilon_{\rm TLD}$) and FOD
  ($\epsilon_{\rm FOD}$) schemes.}  \,
\begin{ruledtabular}
\begin{tabular}{ccccc}
\mbox{H}&\mbox{$\epsilon_{\rm th}= 2 - 2H$}&\mbox{$\epsilon_{\rm X}$}&\mbox{$\epsilon_{\rm TLD}$}&\mbox{$\epsilon_{\rm FOD}$}\\
\hline
\mbox{0.6}&\mbox{0.8}&\mbox{0.838   $\pm$ 0.008}&\mbox{0.830 $\pm$ 0.008}&\mbox{0.834 $\pm$ 0.011}\\
\mbox{0.7}&\mbox{0.6}& \mbox{0.626   $\pm$ 0.004} & \mbox{0.628 $\pm$ 0.004} & \mbox{0.626 $\pm$ 0.004} \\
 \mbox{0.9} &\mbox{0.2} & \mbox{0.206  $\pm$ 0.001} & \mbox{0.225 $\pm$ 0.002} & \mbox{0.211 $\pm$ 0.001} \\
\end{tabular}
\end{ruledtabular}
\end{table}
Pursuing this analysis further, we consider the dynamics of the
processes under study for three other values of $H > 1/2$.  In Table
\ref{Table} we summarize the results obtained by fitting the numerical
data obtained by simulations of the parental process $X(t)$ and of its
two discrete-space and -time counterparts by the power-law function of
the form
\begin{equation}
\label{fit}
\tilde{C}(d\tau) = \frac{1}{(d\tau)^{\epsilon}} \,.
\end{equation}
We observe perfect agreement, in the limits of the statistical error,
between the value of the exponent $\epsilon_{\rm th} = 2 - 2 H$
characteristic of the super-diffusive fBm, the exponent
$\epsilon_{\rm  X}$ deduced for the
numerically generated trajectories $X(t)$, as
well as for the two discrete-space and -time processes obtained within
the TLD ($\epsilon_{\rm TLD}$) and FOD ($\epsilon_{\rm FOD}$)
schemes. This signifies that the TLD and FOD processes can be indeed
considered as the discrete time lattice counterparts of the standard
continuous-space and -time fBm.

The method can be easily applied to $D$ dimensional systems with $D>1$
by using an independent walk for each dimension. 

\begin{figure}
    \centering
    \includegraphics[width=8cm]{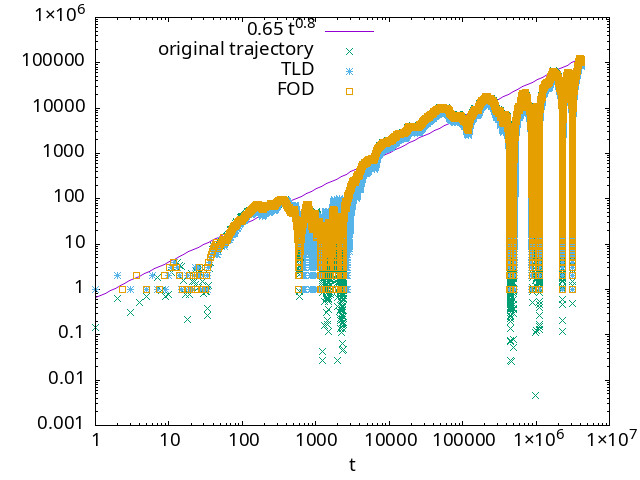}
    \includegraphics[width=8cm]{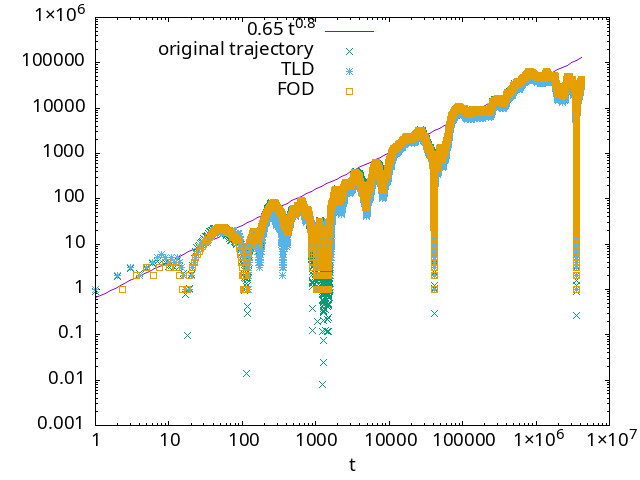}
    \includegraphics[width=8cm]{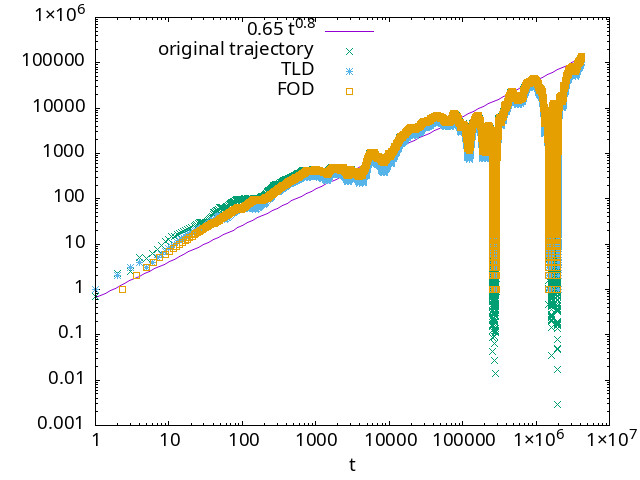}
    \caption{Log-log plot of the time-evolution of absolute values of
      three randomly chosen individual trajectories of a
      super-diffusive fBm with $H= 0.8$. Crosses correspond to a
      numerically generated trajectory $X(t)$ in Eq. \eqref{a} using
      the circulant method, asterisks - to the TLD binarization
      scheme, while the squares show the time-evolution of the
      absolute value of the discrete-space and -time trajectory
      obtained within the FOD approach. Thin magenta line $0.65 \times
      t^{0.8}$ is a guide to the eye.}
    \label{FIG_singletraj_h08}
\end{figure}

Given the very good agreement between the ensemble averaged properties
evidenced in Figs. \ref{FIG_average_h0208} and \ref{fig:2}, we look
next on a more subtle aspect comparing \textit{individual}
trajectories of the process $X(t)$, which corresponds to the
continuous-space and -time fBm obtained using the circulant method,
and its discrete-space and -time counterparts constructed using the
TLD and FOB schemes.  For a super-diffusive fBm with $H= 0.8$, in
Fig. \ref{FIG_singletraj_h08} we present the log-log plot of the
time-evolution of absolute values of three randomly chosen
trajectories $X(t)$ (crosses) together with the trajectories
constructed via the TLD (asterisks) and the FOD (squares) binarization
schemes. In doing so, we observe that there is also a good correspondence
not only between the ensemble averaged properties but also on the
level of individual realizations.

In order to make this statement more precise we quantify the
correlation of the original, continuous trajectory, and the discrete
TLD trajectory, by computing the Pearson coefficient:
\begin{equation}
  r = \frac{\sum_{t=1}^{T} (X_t - \bar{X})(Y_t - \bar{Y})}
  {\sqrt{\sum_{t=1}^{T} (X_t - \bar{X})^2}
    \sqrt{ \sum_{t=1}^{T} (Y_t - \bar{Y})^2} }\,,
\end{equation}
where with $X_t$ and $Y_t$ we denote the two processes we want to
compare. Here, for example, $X_t$ is for the original, space
continuous trajectory, and $Y_t$ for the derived, space discrete TLD
trajectory. For the three cases depicted in
Fig. \ref{FIG_singletraj_h08} we find, respectively, a Pearson
coefficients equal to $.999937$, $.999432$ and $0.999876$. Correlation
of the original trajectory with the generated discrete trajectory is
very high, with a Pearson coefficient very close to one. This implies
that our discretization has kept the original features of the fBM, not
only for average properties, as we have shown, but also for the
individual trajectories, that are linearly dependent from the original
ones. Clearly this is an effect of the main point of this note,
i.e. the fact that the covariance of the increments is preserved by
our binarization.  Our reconstruction has worked very well.

\section{Conclusions}
\label{sec:5}

To conclude, here we have discussed a practically important question
how to define reliably well a lattice version of the super-diffusive
continuous-space and -time fractional Brownian motion - an
experimentally-relevant non-Markovian Gaussian stochastic process with
an everlasting power-law memory on the time-evolution of thermal
noises extending over the entire past. We have put forth two rather
elementary schemes which were shown, through an extensive numerical
analysis, to provide the discrete-space and -time analogues of the
standard continuous-space and -time fBm.  The first approach is based
on a trivial binarization of the increments according to their
signs. In the second approach we have also introduced an
integer valued scale factor taking into account the actual value of
increments. We have shown that for a super-diffusive fBm both
approaches reproduce correctly the power-law decay of the covariance
function of the fBm..

We note that our analysis concerned exclusively the super-diffusive
fBm for which the increments are positively correlated and the
correlations are long-ranged, which circumstance suffices to produce a
super-diffusive motion.  This is not the case for a sub-diffusion, for
which the exponent characterizing the decay of correlations exceeds
unity and hence, the correlations are integrable. In consequence, a
sub-diffusive motion rather takes place due to some intricate
interplay of the sign of increments on successive moves. Finding a
lattice analogue of a sub-diffusion fBm remains a challenging open
problem for further research. We believe that in this way we have
determined an important characterization of fundamental differences of
super-diffusion and sub-diffusion. Controlling super-diffusion is only
a matter of universal quantities. On the contrary our results suggest
that sub-diffusion is also determined by non universal features, local
in time. The small displacements at short times that we are not able
to reproduce in our approach, seem to be crucial for sub-diffusion. It
will be worth it, we believe, to dedicate further work to this issue.

\begin{acknowledgments}
The authors wish to thank Olivier B\'enichou for fruitful discussions.
This project has been supported by funding from the 2021 first FIS
(Fondo Italiano per la Scienza) funding scheme (FIS783 - SMaC -
Statistical Mechanics and Complexity University and Research) from
Italian MUR (Ministry of University and Research).
\end{acknowledgments}

\section*{Data Availability Statement}

Our codes and our data are available from the authors under reasonable
request. 

\section*{References}

\end{document}